\documentclass{easychair}

\usepackage{doc}

\title{Programming Patterns in Dataflow Matrix Machines and
Generalized Recurrent Neural Nets
}

\author{
    Michael Bukatin\inst{1}
\and
    Steve Matthews\inst{2}
\and
   Andrey Radul\inst{3}
}

\institute{
  HERE North America LLC\\
  Burlington, Massachusetts, USA\\ 
  \email{bukatin@cs.brandeis.edu}
\and
  Department of Computer Science\\
  University of Warwick\\
  Coventry, UK\\
  \email{Steve.Matthews@warwick.ac.uk}
\and
   Project Fluid\\
   Cambridge, Massachusetts, USA\\ 
   \email{aor021@gmail.com}
 }

\authorrunning{Bukatin, Matthews and Radul}

\titlerunning{Programming Patterns in Dataflow Matrix Machines}

\begin{document}

\maketitle

\begin{abstract}
Dataflow matrix machines arise naturally in the context of synchronous dataflow programming
with linear streams. They can be viewed as a rather powerful generalization of recurrent neural networks.  Similarly
to recurrent neural networks, large classes of dataflow matrix machines are described by matrices of numbers, and therefore
dataflow matrix machines can be synthesized by computing their matrices. At the same time, the evidence is fairly
strong that dataflow matrix machines have sufficient expressive power to be a convenient general-purpose
programming platform. Because of the network nature of this platform, programming patterns often correspond
to patterns of connectivity in the generalized recurrent neural networks understood as programs. This paper
explores a variety of such programming patterns.
\end{abstract}

\section{Introduction}

There are two lines of thought which lead to dataflow matrix machines (DMMs).

One can start with dataflow programming with {\em linear streams}, i.e. streams admitting linear combinations of several streams~\cite{MBukatinMatthewsLinear}.
The nodes in a dataflow graph compute both non-linear transformations of linear streams and linear combinations of those streams. 
If one adopts the discipline of not allowing to connect nodes computing non-linear transformations of streams directly to each other,
but only allowing to connect them via nodes computing linear combinations of streams, then one can
parametrize large classes of dataflow programs by matrices~\cite{MBukatinMatthewsMatrices}.

Alternatively, one can start with recurrent neural networks (RNNs) and generalize them as follows.
\begin{itemize}
\item Allow neurons to process not just streams of reals, but a diverse collection of linear streams.
\item Allow not only neurons with one linear combination as an input, but also neurons
with two linear combinations as inputs, neurons with three linear
combinations as inputs, etc.\footnote{Neurons without inputs are also allowed and serve as input streams for the network.}
\item Allow neurons to have not just one output, but also two outputs,
three outputs, etc.\footnote{Neurons without outputs are also allowed for side-effects, such as external recording of the results.}
\item Allow neurons to have a rich collection of built-in
stream transformers.
\end{itemize}
One can parametrize the large classes of so generalized
RNNs by their matrices in a manner similar to parametrization of ordinary RNNs by matrices~\cite{MBukatinMatthewsRadulPrelim,MBukatinMatthewsRadulDMM}.

The matrix parametrization of the large classes of the resulting dataflow machines (DMMs)
means that various methods of synthesizing and training RNNs tend to be applicable to DMMs.

At the same time, the power of DMMs makes them a convenient general-purpose
programming platform. In~\cite{MBukatinMatthewsRadulDMM} we started to sketch programming idioms and
constructions for DMMs.

\subsection{Programming Patterns}

The purpose of the present paper is to develop and present a number of such programming
idioms and constructions in sufficient detail to enable the use of DMMs in daily engineering work.

It is natural to think about those idioms and constructions as {\em connectivity patterns} in
our generalized neural networks. Therefore in this paper we speak in terms of {\em programming patterns} and
{\em programming with patterns} in generalized neural networks.

The programming patterns we discuss include

\begin{itemize}
  \item Identity transform and accumulators;
  \item Two inputs and multiplicative masks;
  \item Piecewise bilinear neurons and ReLU;
  \item Reflection facilities: a network can modify its own matrix;
  \item Deep copy of a subnetwork;
  \item Nested deep copy;
  \item Active data: allocating the linked structures in the body of the network.
\end{itemize}

\subsection{Example of a Programming Task}\label{example_task}

The focus of this paper is on using DMMs as a programming platform for manually written code rather than
on automatically learning the DMM programs, which should be the subject of a separate paper.
We use the following programming task as an example.

\paragraph{Find whether a string has duplicate characters.}
This is a variant of the first coding interview problem in~\cite{GMcDowell}. It is a simple coding problem with some subtle aspects.\footnote{By ``duplicate characters" we mean characters occurring more than once in the string.}
Hence we find it attractive to take this coding problem as a starting point of our exploration.\footnote{The detailed solution is in Section~\ref{software_example}. The comparison of the solution using DMMs versus solutions using traditional RNNs is in Section~\ref{related_work}.}

\paragraph{Characters as vectors.} The standard representation of characters in neural nets is by vectors with the dimension
of the alphabet in question via the ``1-of-$N$" encoding~\cite{AKarpathy}. A character is represented by a vector with 1 at the coordinate corresponding
to this character, and with zeros at the rest of the coordinates.
Our architecture is friendly to the use
of sparse vectors when desired, which is particularly valuable for large alphabets such
as the collection of Unicode characters.

\paragraph{Linear streams.} 
For this programming task we use two kinds of linear streams: streams of numbers (we call them {\em scalars}) 
and streams of vectors
representing characters in the ``1-in-$N$" encoding (we call those {\em c-vectors}).

\paragraph{Input neuron.}
Let the string be emitted by an input neuron, which has no inputs of its own, emits one c-vector representing a character
per clock tick, and emits the c-vector representing the end-of-string character (EOS) at the end of the string. EOS is
considered to be a particular letter of the alphabet.

We allow the input neuron to emit  characters at a slower rate than clock ticks. If there is no character to
emit at a given clock tick, then the zero c-vector is emitted on that clock tick.

\subsection{Programming Patterns: More Details}

Impose the precision requirement on our number system that it has exact 0 and exact 1,
that multiplication by exact 0 yields exact 0, and that addition of exact 0 to $x$ and
multiplication of $x$ by exact 1 leave $x$ exactly unchanged.

\subsubsection{Identity Transform and Accumulators}

It is extremely useful to allow neurons with the built-in transform being
identity transform from the input stream to the output stream of the same kind. Usually one does not
include this type of neurons in conventional neural nets. However, having neurons
equipped with the identity transform yields a lot of expressive power. Consider a neuron
equipped with identity transform from the input $x$ to the output $y$. 
Just like any input, the input $x$ is a linear combination of various neuron outputs.
Set the coefficient of dependency of $x$ from $y$ as 1. Then the neuron in
question works as a memory accumulator. It accumulates the contributions of
all other neuron outputs connected to $x$ with non-zero weights. 

If one sets the coefficient of dependency of $x$ from $y$ to be $0 < \alpha < 1$,
then the neuron in question is a leaky accumulator: during each cycle of the
network the previously accumulated value is multiplied by $\alpha$.

Using accumulators and leaky accumulators, one can easily implement
integrate-and-fire and leaky integrate-and-fire schemas, and have them not
as a separate network of spiking neurons, but as a part of regular
conventional recurrent neural network, where some neurons happen to
spike.

If one sets the coefficient of dependency of $x$ from $y$ to be -1, then
one gets an oscillator, with input from other neurons additively modulating
the oscillations.

\paragraph{Accumulator of c-vectors.} While solving the problem of detecting
duplicate characters in a string, we need to accumulate the
sum of c-vectors corresponding to the characters of the input string. When the value of any coordinate of
the accumulated sum exceeds 1, this would indicate the presence of duplicate characters. Hence we
connect the output of the input neuron to the input of the accumulator of c-vectors with the weight 1 (Section~\ref{software_example}).

\subsubsection{Two Inputs and Multiplicative Masks}

The idea of allowing neurons with two and more linear combinations as inputs
has been proposed several times. In particular, it is well-known that
multiple inputs allow us to have multiplicative neurons which compute 
polynomials and that a lot of convenience and pragmatic engineering
power can be derived from it (see e.g. Section 4.6 of~\cite{JPollack} and
references therein).

A particularly important case is when one of the inputs (for a
linear stream of numbers) is used as a scalar multiplicative mask. i.e.
the neuron outputs are multiplied by the current value of that
input.

Setting the multiplicative mask to zero allows one to suppress and turn off
parts of the network in a dynamic manner, with patterns of activity suppression
changing from moment to moment.

In particular, multiplicative masks can be used to express various conditional
constructions, and to redirect flows of data in the network.

Setting multiplicative masks to zero allows one to express various kinds of
precise orchestration. For example, one often wants to express layers
typical for deep learning models within a recurrent neural network.
This is done by specifying the appropriate connectivity patterns
in the connectivity matrix of the network. However, in addition
one often wants to prevents layers from firing all at once, but
would like them to fire one after another. Multiplicative masks are
ideal for this kind of orchestration.

At the same time, generally varying the values of multiplicative masks
often allows to express linear combinations varying in time without modifying
the network matrix itself.

\paragraph{Multiplicative mask for identity transform.}
Consider the type of an accumulator neuron having one vector input $x$ and one scalar input $a$, and
one vector output $y$ which gets the value $a*x$. The output $y$ is connected to the input $x$ by the weight 1, 
which makes this neuron to actually function as accumulator of its vector inputs, when $a$ is set to 1.
Here $a$ is used as a {\em multiplicative mask} for the accumulator in question.
It is normally set to 1, but when there is a need to reset the accumulator, this mask is set to 0.
Setting the mask to $0 < \alpha < 1$ makes the neuron to function as a leaky
accumulator, and setting it to -1 makes the neuron to function as an oscillator.

\paragraph{Piecewise bilinear neurons and ReLU.}
The rectifier linear unit (ReLU), with the piecewise linear activation function $f(x) = \mbox{max}(0,x)$, 
is a particularly popular type of neurons in the neural net community in recent years~\cite{YLeCunBengioHinton}.

What should work particularly well in light of the discussion in the current subsection is the piecewise bilinear rectifier,
with the activation function $g(x,y) = \mbox{max}(0,x)*\mbox{max}(0,y)$.

 With this formula, both inputs serve as multiplicative masks for each other,
and this is being combined with the well known power of ReLU neurons,
and well known power of bilinear neurons.

\subsubsection{Reflection Facilities: a Network Can Modify Its Own Matrix}\label{reflection_facilities}

In this approach one always thinks that there is a countable number
of neurons of each predefined type, and therefore any DMM over a
particular {\em signature} specifying the kinds of available linear
streams and the types of available neurons is determined by a countable
connectivity matrix. We impose a condition that only a finite number
of elements of this matrix is non-zero at any given moment of time.

One of the key achievements of~\cite{MBukatinMatthewsRadulDMM} is that among the linear streams
in question it allowed the streams of matrices shaped like the matrices controlling
the network in question, and also the streams of rows and the streams of
columns of such matrices. 

A dedicated neuron, {\tt Self}, equipped with the identity transform of the stream of matrices
controlling the network in question is used as an accumulator. The input of {\tt Self}
adds together the updates to the network matrix made by other neurons of the network,
and the action of {\tt Self} makes those updates available for use when the
inputs of the neurons are recomputed from the outputs of the neurons during
the next cycle.

Therefore, {\tt Self} enables other neurons to both use the network matrix as one of the
inputs, and to modify the network matrix by supplying additive updates.

So in this approach, the network can meaningfully modify itself during its functioning.
Both the particular values of non-zero weights, and the sparsity structure,
i.e. the difference between non-zero and zero weights which defines the
network topology, are subject to such modification.

\subsubsection{Deep Copy of a Subnetwork}

One important construction is deep copy of a subgraph. 
The structure of the incoming connections external to the subgraph in question is
preserved for the resulting new subgraph. The outgoing connections for the
resulting new subgraph might be set to zero (similarly to~\cite{MBukatinMatthews}), or
copied from the outgoing connections of the original subgraph, or some of the weight 
of the outgoing connections of the original subgraph can be transferred to the outgoing connections of
the resulting new subgraph.

In~\cite{MBukatinMatthews}, we described an algorithm for this operation as a graph algorithm.
Here we describe this operation in terms of matrices (Section~\ref{deep_copy_via_matrix}), which sheds additional
light on its nature.

\paragraph{Nested deep copy.}
The deep copy operation can be repeated a number of times, creating a pattern of
copies of a particular subgraph.

It can also be applied in a nested fashion, thus creating intricate ``pseudo-fractal"
connectivity patterns.

Because we have reflection facilities, neurons of the network itself are capable of
creating deep copies of the network's subgraphs. Hence the pattern creation in the
network can be controlled by the network itself.

\paragraph{Silent and active parts of the network.} The standard approach of DMMs is that
there is a countable number of neurons of each type, but only finite number of nonzero
elements in the countable-sized network matrix, and all neurons which have some nonzero
connectivity are active, while other neurons are silent and not present in memory.

However, it is convenient to be able to have silent parts of the network with non-zero 
connectivity weights. E.g. one might accumulate a library of connectivity patterns and build
a network from those patterns via the deep copy facilities. It does not make sense for the library instances to also function
and consume CPU.

It is also often the case that one builds a network gradually. In some contexts, one wants to
gradually build or change a network, while this network is functioning (the ability to do so is a powerful and
unusual feature of this approach). But in a number of contexts, it is preferable to build a silent network,
then to deep copy it into the active area, so that it gets initialized and starts functioning all at once.

\subsubsection{Active Data: Allocating the Linked Structures in the Body of the Network}

The structure of weighted connections between neuron outputs and neuron inputs can be used
to represent various linked data structures, such as variants of lists, trees, and graphs.
The linked structures in question can be located both in silent and active parts of the network.

A particularly interesting case is when the linked structures are located in the active part of
the network. Then there are various ways to use the functioning of the neurons involved in such
data structures. In this case we can speak about {\em active data}. Because the rows (or, less frequently, the columns) of the network matrix can be used
to represent the links, and because we allow types of neurons which take and emit streams of
appropriately shaped rows (or columns), one can use those streams of rows (or columns) to represent the linked structures
even without using the rows (or columns) of the network matrix itself.

In particular, the accumulator metaphor is often useful to carry the payload of a
node. The accumulator metaphor is also useful to hold the row (or the column) representing the links from that node
in those situations where we opt not to use rows (or columns) of the network matrix itself
for that purpose.

\subsection{Design Philosophy: Rich Signatures}

The various universality results notwithstanding, when one tries to use recurrent neural networks
as a general-purpose programming platform, one finds that austere selection of linear streams
(only streams of numbers with element-wise linear combinations) and typically very limited
selection of types of neurons interfere with the pragmatic needs of a software engineer.

Here we follow a different design philosophy. When there seems to be a need in a new kind
of linear streams or a new type of neurons, we simply add those new kinds of streams and
new types of neurons to the signature of our formalism. Practically speaking, we envision that
for industrial uses one would have dozens of kinds of linear streams and dozens of types of
built-in neurons in the DMM signature.

Our formalism for defining new types of neurons is powerful enough to give us an option to take any subnetwork
and make a new type of neurons, such that a single neuron of this new type is equivalent to
the subnetwork in question.

In turn, the universality results would in many cases allow one to approximate new types of neurons
by subnetworks of existing neurons with higher clock speed, but only at great computational
cost.

\subsection{Structure of the Paper}

Section~\ref{related_work} discusses related work and briefly introduces the necessary notions
from~\cite{MBukatinMatthewsRadulDMM}. It also discusses the differences
between programming in DMMs and programming in the traditional RNNs.
Section~\ref{software_example} contains a complete implementation of a detector of
duplicate characters as a DMM. Section~\ref{deep_copy_via_matrix} describes details of a
few variants of the operation performing deep copy of a subnetwork in terms of
transformations of a network matrix.

 In Conclusion, we note
that the diversity of available programming patterns means that a considerable
variety of programming styles and paradigms should be possible for DMMs.

\section{Related Work}\label{related_work}

This work has its roots in two fields, dataflow programming with streams of
continuous data (such as, for example, LabVIEW and Pure Data dataflow programming languages~\cite{WJohnstonHannaMillar,AFarnell})
and  recurrent neural networks.

Turing universality of recurrent neural networks seems to be known for at least 30 years~\cite{JPollack,HSiegelmannSontag}.
Nevertheless, and notwithstanding the remarkable progress in the methods of training and in applications of recurrent
neural networks, they have not become a convenient general-purpose programming platform.

The example of a simple detector of duplicate characters in Section~\ref{software_example} of the present paper sheds some light on the
underlying reasons for that. The dataflow matrix machine implementing this example in the present paper consists of 9 neurons and 10 nonzero scalar connection weights.

If one would implement the same algorithm in traditional recurrent neural nets oriented towards
working with streams of numbers, then for a typical small alphabet of a few dozen allowed
characters one would need a few hundred neurons and a few hundred nonzero scalar connection weights.
While this is quite manageable, the inconvenience is obvious.

The situation becomes much more serious when one wants to solve the same problem for a large set
of characters, such as, for example, Unicode. The DMM architecture is quite friendly to sparse high-dimensional vectors, so
the same compact program would work, the only aspect which would change is the definition and implementation
of the stream of c-vectors. On the other hand, the traditional recurrent neural net architecture oriented towards
working with streams of numbers is not friendly towards naively handling sparse high-dimensional arrays. Given that
the complete Unicode set is well above 100,000 characters, the task of making a traditional recurrent network which
would solve this simple problem seems to be quite non-trivial (at the very least, it looks like one would need an algorithmic breakthrough
to find an exact solution of reasonable size and efficiency).

\subsection{Timeline of Dataflow Matrix Machines}

The architecture oriented towards working with a variety of linear streams such as probabilistic sampling and generalized animations
was proposed in~\cite{MBukatinMatthewsLinear}. The discipline of bipartite graphs allowing to represent
large classes of dataflow programs working with linear streams by matrices and first variants of higher-order constructions allowing
to continuously transform a dataflow program while it is running emerged in~\cite{MBukatinMatthewsMatrices}.
While in dataflow programming languages oriented towards work with streams of
continuous data such as LabVIEW and Pure Data the programs were represented by
discrete objects, dataflow matrix machines were themselves continuous.

An observation that dataflow matrix machines might be viewed as
generalized recurrent neural nets was made in~\cite{MBukatinMatthewsRadulPrelim}.
The systematic development of dataflow matrix machines as a programming platform was started in~\cite{MBukatinMatthewsRadulDMM}.
A particularly important new development in~\cite{MBukatinMatthewsRadulDMM} was introduction of high-level
reflection facilities via streams of appropriately shaped matrices and their rows and columns (see Section~\ref{reflection_facilities} of the present paper).
A sketch of a programming language to describe DMMs was also started in~\cite{MBukatinMatthewsRadulDMM}.
 
That language included declarations of kinds of linear streams, \texttt{\#kind}, and declarations of types of neurons, \texttt{\#newcelltype}.
The purpose of those declarations was to make names of new kinds of linear streams and types of neurons to be
available to the program, and to link those names to the implementations of the corresponding linear streams and of
the corresponding built-in transforms (associated with types of neurons) in the underlying conventional programming language.

That language also included declarations of particular neurons, \texttt{\#neuron}, solely for the purpose of picking a neuron of a specific type
with zero connectivity for all its inputs and outputs and giving names to this neuron and to its inputs and outputs.

The only functional operator affecting the network behavior which was introduced in~\cite{MBukatinMatthewsRadulDMM} was an additive operator
updating matrix weights. It was introduced in several forms. The most generic of those forms was

\texttt{\#updateweights <FiniteRowMask 1> += <ColumnMask> * <FiniteRowMask 2>;}

which on the matrix level was acting as $a_{ij} := a_{ij} + \gamma_i * \alpha_j * \sum_k \beta_k a_{kj}$, where
\texttt{<ColumnMask>} was vector $\alpha$, \texttt{<FiniteRowMask 2>} was vector $\beta$, and the left-hand-side 
\texttt{<FiniteRowMask 1>} was vector $\gamma$. In our present example program only the most
simple form of this operator adding 1 to a specific weight from a particular output to a particular input is used:

\texttt{\#updateweights <IdInputStream> += <IdOutputStream>; }

Finally, the principle of having sufficiently powerful neuron types to express the network updates available in
the language describing the DMMs was formulated in~\cite{MBukatinMatthewsRadulDMM}. This principle means
that a network is supposed to have full access to all our language facilities and can use those facilities to modify itself.

\section{A Version of Duplicate Characters Detector}\label{software_example}

Let's sketch a DMM solving the coding problem described in Section~\ref{example_task}.
We have two kinds of linear streams, scalars and c-vectors: 
\texttt{\#kind real; \#kind c-vector;}

\subsection{Input Accumulator Circuit}

The type of the input neuron is

\texttt{\#newcelltype input-string \#output c-vector:emit;}\newline
The type of the accumulators (the  identity transforms) of c-vectors is

\texttt{\#newcelltype id-c-vector \#input c-vector:in \#output c-vector:out;} \newline
Pick a neuron of each of these types with zero connectivity patterns, and give names
to those neurons, and to their input and output streams.

\texttt{\#neuron  input-string:input-data emit:emit-c-vector = \#transformof \#dummy;}

\texttt{\#neuron id-c-vector:accumulator out:accum-pass-through =}

\texttt{\ \ \ \ \#transformof  in:collect-sum;}\newline
Link the accumulator output to the accumulator input with weight 1:

\texttt{\#updateweights collect-sum += accum-pass-through;}\newline
Link the emitter of the input to the input of the accumulator with weight 1:

\texttt{\#updateweights collect-sum += emit-c-vector;}

\subsection{Circuit Detecting Duplicate Characters}

The type of a neuron computing the maximal absolute value of a c-vector coordinate:

\texttt{\#newcelltype max-norm-of-c-vector \#input c-vector:in \#output real:max-norm;}\newline
Pick a neuron of this type with zero connectivity patterns, and give names to
this neuron and to its input and output streams:

\texttt{\#neuron max-norm-of-c-vector:eval-max-char-count max-norm:max-char-count =}

\texttt{\ \ \ \ \#transformof in:c-vector-to-measure;}\newline
Link the accumulator output to the input of this neuron with weight 1:

\texttt{\#updateweights c-vector-to-measure += accum-pass-through;}\newline
Next, we need scalar constant 1:

\texttt{\#newcelltype input-real \#output real:emit;}

\texttt{\#neuron input-real:const-1-stream emit:const-1 = \#transformof \#dummy;}\newline
Next, we need a neuron comparing two scalars. It is convenient to equip neurons which compute
conditions with two complementary output scalars:

\texttt{\#newcelltype greater-than}

\texttt{\ \ \ \ \#input real:scalar-to-be-greater \#input real:scalar-to-be-smaller}

\texttt{\ \ \ \ \#output real:true-channel \#output real:false-channel;}

\texttt{\#neuron greater-than:test-max-norm-greater-than-1}

\texttt{\ \ \ \ true-channel:duplicate-detected false-channel:ignore = \#transformof}

\texttt{\ \ \ \ scalar-to-be-greater:input-norm scalar-to-be-smaller:input-const-1;}\newline
Now link the norm and the const 1 to the respective inputs of this neuron with weight 1:

\texttt{\#updateweights  input-norm += max-char-count;}

\texttt{\#updateweights input-const-1 += const-1;}

\subsection{Circuit Detecting End-of-string}

The dot product of c-vectors is a good way to
detect presence of a particular character:

\texttt{\#newcelltype dot-product-of-c-vectors}

\texttt{\ \ \ \ \#input c-vector:in-1 \#input c-vector:in-2 \#output real:dot-product;}

\texttt{\#neuron dot-product-of-c-vectors:end-of-string-detector}

\texttt{\ \ \ \ dot-product:end-of-string-predicate =}

\texttt{\ \ \ \ \#transformof in-1:c-vector-to-test in-2:const-char;}\newline

Now we have a choice to detect this end of string at the level of \texttt{input-data} neuron,
or at the level of \texttt{accumulator} neuron. Let's for the time being detect it at the level of
the accumulator:

\texttt{\#updateweights c-vector-to-test += accum-pass-through;}\newline
The end-of-string constant is similar to \texttt{input-string}, but with a different built-in
transform:

\texttt{\#newcelltype end-of-string-const \#output c-vector:emit;}

\texttt{\#neuron  end-of-string-const:end-of-string-const-stream}

\texttt{\ \ \ \ emit:const-eos = \#transformof \#dummy;}

\texttt{\#updateweights const-char += const-eos;}\newline

Now we need to use another neuron comparing two scalars. The nice feature
is that const zero can be omitted (one has zero linear combination on an input
by default), so no explicit link to \texttt{input-const-0} is required:

\texttt{\#neuron greater-than:test-eos-presence-predicate}

\texttt{\ \ \ \ true-channel:eos-detected false-channel:ignore =}

\texttt{\ \ \ \ \#transformof scalar-to-greater:input-eos-presence-predicate}

\texttt{\ \ \ \ scalar-to-be-smaller:input-const-0;}

\texttt{\#updateweights input-eos-presence-predicate += end-of-string-predicate;}\newline

We are almost done. If we run the program, we shall eventually get 1 on the
\texttt{duplicate-detected} stream or on the \texttt{eos-detected} stream,
and then we'll know whether the input string has a duplicate character or not.

\subsection{Output neuron with side effect}\label{output_side_effect}

Now we can optionally add the specialized neuron for recording the output and
stopping the network as a side effect. It does not need to have outputs visible to
the network, so it is a ``dual" to an input neuron in that it has inputs, but zero outputs:

\texttt{\#newcelltype record-answer-and-stop-the-network}

\texttt{\ \ \ \ \#input real:positive-answer \#input real:negative-answer}

\texttt{\#neuron  record-answer-and-stop-the-network:output-control \#dummy =}

\texttt{\ \ \ \ \#transformof positive-answer:duplicates-present}

\texttt{\ \ \ \ negative-answer:duplicates-absent;} 

\texttt{\#updateweights duplicates-present += duplicate-detected;}

\texttt{\#updateweights duplicates-absent += eos-detected;}

\subsection{Real-time aspects}

In the synchronous dataflow precise timing of events is often important.
Let's take note of how many clock cycles does it take for a new c-vector emitted by
\texttt{input-data} to propagate to various neurons in our first version of
a duplicate character detector:

\begin{itemize}
  \item \texttt{accumulator} neuron: 1 clock cycle;
  \item \texttt{eval-max-char-count} and \texttt{end-of-string-detector} neurons: 2 clock cycles;
  \item \texttt{test-max-norm-greater-than-1} and \texttt{test-eos-presence-predicate}: 3 clock cycles;
  \item \texttt{output-control} neuron: 4 clock cycles.
\end{itemize}

It is important that the propagation delay to \texttt{test-max-norm-greater-than-1} is not larger than
the propagation delay to \texttt{test-eos-presence-predicate}, otherwise a wrong answer might be
recorded. In particular, one can prove that \texttt{duplicates-present} and \texttt{duplicates-absent}
are never triggered simultaneously when those propagation delays are equal.

While programming in this style, one needs to keep those timing issues in mind. Sometimes
one has to insert extra delays along certain signal propagation paths or to refrain from optimizing
certain propagation paths to ensure the correct timing conditions.

\section{Deep Copy of a Subnetwork via Network Matrix}\label{deep_copy_via_matrix}

We'll need new syntax to assemble a group of neurons into a subgraph, and to access the neurons
and their inputs and outputs from the deep copy of that subgraph.

\texttt{\#subgraph name-of-the-new-subgraph = \#cells neuron-name-1 \dots neuron-name-N;}

\texttt{\#new-copy name-of-the-new-subgraph = \#deepcopyof old-subgraph;}\newline
There are variants of deep copy operation, for example:

\begin{enumerate}
  \item deep copy a subgraph, but omit the external connectivity;
  \item deep copy a subgraph, copy the external incoming connections of the original subgraph into the corresponding nodes of the new copy of that subgraph, but omit the external outgoing connections (this variant was considered in~\cite{MBukatinMatthews});
  \item deep copy a subgraph and copy the external incoming and outgoing connections;
  \item deep copy a subgraph and copy the external incoming connections and distribute some of the weight of the outgoing connections of the original graph into the outgoing connections of the new copy.
\end{enumerate}

One needs variations of the \texttt{\#new-copy} syntax to reflect these possibilities.

Names of subgraphs form namespaces. Neurons and their inputs and outputs can be referenced via those namespaces in the language describing the DMMs.

Let's look at the variants of \texttt{\#new-copy} on the level of network matrix.

\begin{figure}[h]
   \begin{picture}(400,160)
   \put(30,130){\line(1,0){340}}
   \put(30,100){\line(1,0){340}}
   \put(60,155){\line(0,-1){150}}
   \put(150,155){\line(0,-1){150}}
   \put(65,112){Original subgraph}
   \multiput(30,60)(21,0){17}{\line(1,0){5}}
   \multiput(30,30)(21,0){17}{\line(1,0){5}}
   \multiput(250,155)(0,-21){8}{\line(0,-1){6}}
   \multiput(340,155)(0,-21){8}{\line(0,-1){6}}
   \put(265,42){The new copy}
   \put(280,112){Zero}
   \put(90,42){Zero}
   \end{picture}
   \caption{Deep copy of a subnetwork in terms of connectivity matrix}
   \label{fig_deep_copy}
\end{figure}
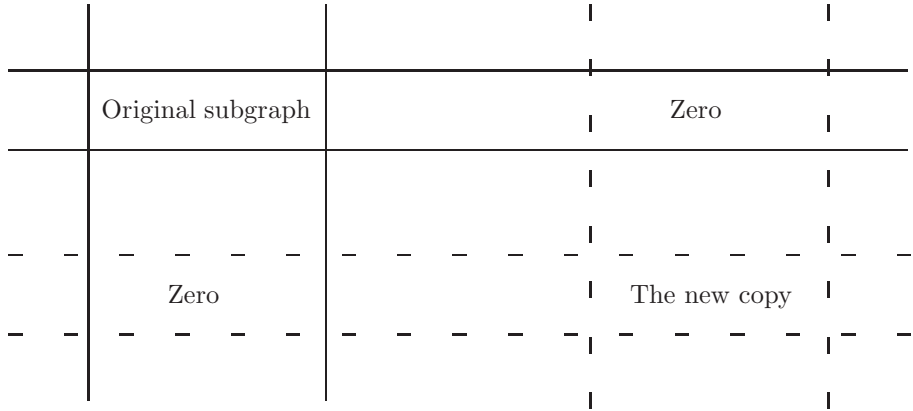

Let's imagine for the duration of this section that indices of rows and columns form a traditional one-dimensional structure,
and that the rows corresponding to the inputs of the original subgraph are grouped together in the space of row indices,
and that the columns corresponding to the outputs of the original subgraph are grouped together in the space of column indices.
Then the original subgraph is represented by solid lines in the Figure~\ref{fig_deep_copy}.

The rows are incoming connections of the subgraph, and the columns are outgoing connections of the subgraph, with
their intersection representing internal connectivity of the subgraph.

To create the new copy, find a similarly shaped patch of unused neurons. In terms of the matrix, pick the appropriate
range of rows and columns such that all those rows and columns are currently zeros (dashed lines in the Figure~\ref{fig_deep_copy}).

To implement variant 1, simply copy the matrix block marked by the words ``Original subgraph" to the matrix block
marked by the words ``The new copy".

To implement variant 2, copy the rows of the original subgraph into the rows of the new copy, but place all zeros in
the block where the copy of the internal structure of the original subgraph would be (the word ``Zero" in the lower left
corner of  Figure~\ref{fig_deep_copy}), and instead copy the matrix block marked by the words ``Original subgraph" to the matrix block
marked by the words ``The new copy".

To implement variant 3, a similar operation should also be done with the columns.

To implement variant 4, one would implement variant 3 and then multiply the parts of new columns situated outside ``The new copy"
block by $\alpha$, and multiply the parts of the original columns situated outside ``The original subgraph" block by $1-\alpha$.

\subsection{Higher-order Neurons for Deep Copy}

Section 3.7 of~\cite{MBukatinMatthewsRadulDMM} describes higher-order neurons for the \texttt{\#updateweights} operation.
Similarly, in order to conform to our principle that the language facilities should be available from within the network via
higher-order neurons, we describe higher-order neurons for the \texttt{\#new-copy} operation. These higher-order
neurons tend to be rather complex. Speaking in terms of Section 3.7 of~\cite{MBukatinMatthewsRadulDMM} , one would need
the matrix as an input, two more inputs (a ``row mask" and a ``column mask") to describe the subgraph to copy,
and a scalar multiplicative mask to control the firing. There would be several outputs, one is an additive contribution to the matrix
similar to  Section 3.7 of~\cite{MBukatinMatthewsRadulDMM}. The other two outputs are a ``row mask" and a ``column mask"
expressing the new subgraph.

Subgraphs to be copied can contain such higher-order neurons, and it is quite legal for a higher-order neuron of this kind
to copy a subgraph containing this neuron itself.

\subsection{Gradual Updates and Gradual Creation of a New Subgraph Copy}

It is often desirable to have continuous updates to weights rather than abrupt updates. Since we presently consider
discrete time, this would means smaller updates of weights performed during several consequent clock ticks.
For the neuron implementing \texttt{\#updateweights} operation the most natural way to accomplish this
is via small non-zero values $\alpha$ for its input scalar multiplicative mask.

For the neuron implementing \texttt{\#new-copy} the situation is more delicate, because this neuron causes
allocation of a new subgraph in the unused address space of the network (the space containing neurons with
zero input and output connectivity). We need to equip such a neuron with memory to make sure that
this allocation happens only when the multiplicative mask changes from zero to non-zero, but that otherwise
the weight changes are applied to the existing new copy. The easiest way to achieve that is to create two extra vector
inputs to store the output ``row mask" and ``column mask" via the accumulator metaphor.

\section{Conclusion}

In this paper we developed a number of powerful programming patterns for dataflow matrix machines.

An example of a DMM for a simple coding problem of
detecting characters occurring multiple times in a string is very compact,
unlike the counterpart of the same algorithm which could be written in traditional
recurrent neural networks working with streams of numbers. The difference is particularly striking for
large alphabets necessitating the use of sparse arrays.

\subsection{Design Philosophy: Pluralism of Programming Styles}

Traditional programming architectures with programs expressed as discrete objects support a large
variety of programming styles: imperative, object-oriented, functional, logical, dataflow, and many others.

We expect that the presently emerging new realm of programs expressed as continuous objects, specifically as matrices of numbers,
will also support a diversity of programming styles. Some of those programming styles might be fairly conservative modifications
of programming styles available in the realm of discrete programs, while other styles might be entirely novel.
The emerging architecture of continuous programs seems to
be sufficiently unique to allow creation of entirely new programming styles.\footnote{{\bf August 2018 note:} Reference paper on dataflow matrix machines 
(DMMs) research in 2016-2017 is \url{https://arxiv.org/abs/1712.07447}. The most important innovation introduced there is 
spaces of {\bf V-values} (vector-like elements based on nested maps, Section 3) which enable DMMs with {\bf variadic neurons} (Section~4) and conveniently represent conventional data structures (Section 6.3). See Section 6.4 of that paper for discussion of programming with streams of V-values.}

%------------------------------------------------------------------------------
%\section{Acknowledgments}

%THIS SECTION WILL BE WRITTEN LATER.

\label{sect:bib}
\bibliographystyle{plain}
%\bibliographystyle{alpha}
%\bibliographystyle{unsrt}
%\bibliographystyle{abbrv}
%\bibliography{dmm_patterns_v2}

\end{document}